\documentclass[prb,nofootinbib,twocolumn,showpacs,superscriptaddress,floatfix]{revtex4}

\usepackage{graphicx}
\usepackage{amsmath, amssymb, amsfonts}

\newcommand{\lsco}{La$_{1.83}$Sr$_{0.17}$CuO$_4$}
\newcommand{\lscohead}{L\lowercase{a}$_{\mathbf{1.83}}$S\lowercase{r}$_{\mathbf{0.17}}$C\lowercase{u}O$_{\mathbf{4}}$}

\newcommand{\lscox}{La$_{2-x}$Sr$_{x}$CuO$_4$}
\newcommand{\ybco}{YBa$_{2}$Cu$_{3}$O$_{7-\delta}$}

\newcommand{\bscco}{Bi$_2$Sr$_2$CaCu$_2$O$_{8+\delta}$}

\begin{document}
\title{Study of the mixed state of \lscohead\ by means of muon-spin rotation and magnetization experiments in a low magnetic field}

\author{B.~M. Wojek}
\altaffiliation{Present address: Material Physics, KTH Royal Institute of Technology, Isafjordsgatan 22, 16440 Kista, Sweden}
\email{basti@kth.se}
\affiliation{Labor f{\"u}r Myonspinspektroskopie, Paul Scherrer Institut, 5232 Villigen PSI, Switzerland}
\affiliation{Physik-Institut der Universit{\"a}t Z{\"u}rich, Winterthurerstrasse 190, 8057 Z{\"u}rich, Switzerland}
\author{S.~Weyeneth}
\affiliation{Physik-Institut der Universit{\"a}t Z{\"u}rich, Winterthurerstrasse 190, 8057 Z{\"u}rich, Switzerland}
\author{S.~Bosma}
\affiliation{Physik-Institut der Universit{\"a}t Z{\"u}rich, Winterthurerstrasse 190, 8057 Z{\"u}rich, Switzerland}
\author{E.~Pomjakushina}
\affiliation{Laboratory for Developments and Methods, Paul Scherrer Institut, 5232 Villigen PSI, Switzerland}
\author{R.~Pu\'{z}niak}
\affiliation{Institute of Physics, Polish Academy of Sciences, Aleja Lotnik\'{o}w 32/46, 02-668 Warsaw, Poland}

\date{\today}

\begin{abstract}
Muon-spin rotation ($\mu$SR) experiments are often used to study the magnetic field distribution in type-II superconductors in the vortex state. Based on the determination of the magnetic penetration depth it is frequently speculated---also controversially---about the order-parameter symmetry of the studied superconductors. This article reports on a combined $\mu$SR and magnetization study of the mixed state in the cuprate high-temperature superconductor \lsco\ in a low magnetic field of $20$~mT applied along the $c$ axis of a single crystal. The macroscopic magnetization measurements reveal substantial differences for various cooling procedures. Yet, indicated changes in the vortex dynamics between different temperature regions as well as the results of the microscopic $\mu$SR experiments are virtually independent of the employed cooling cycles. Additionally, it is found that the mean magnetic flux density, locally probed by the muons, strongly increases at low temperatures. This can possibly be explained by a non-random sampling of the spatial field distribution of the vortex lattice in this cuprate superconductor caused by intensified vortex pinning.
\end{abstract}
\maketitle

\section{Introduction}

The muon-spin rotation ($\mu$SR) technique is a powerful tool to study the local magnetic field distribution in solids.\cite{Blundell1999} It is also successfully employed to gain valuable information on type-II superconductors by probing the magnetic field distribution $P(B)$ generated by a vortex lattice (for a review mostly regarding cuprate high-temperature superconductors, see, e.\,g. Ref.~\onlinecite{Sonier2000}). If the vortex-lattice configuration is ordered and static, from the obtained $P(B)$ the characteristic length scales of the superconductor like the magnetic penetration depth $\lambda$ or with reservations the Ginzburg-Landau coherence length $\xi$ can be extracted in a reliable way. However, the vortex arrangement in cuprate superconductors is commonly not an ideal static two-dimensional hexagonal flux-line lattice. Individual vortices are always subject to displacements due to pinning, e.\,g. at grain boundaries or lattice defects. Also either thermal or quantum fluctuations introduce dynamics and may cause a reordering and relaxation of a non-equilibrium vortex lattice. Reviews on this subject can be found, e.\,g. in Refs.~\onlinecite{Blatter1994} and~\onlinecite{Brandt1995}. The presence of those imperfections changes the $\mu$SR field distributions: while weak random pinning leads to a symmetric broadening of $P(B)$,\cite{Brandt1988} thermal fluctuations might even change the observed asymmetry of the lineshape as seen in the example of the vortex-lattice melting in the \bscco\ compound.\cite{Lee1993} Moreover, the lineshape asymmetry has also been found to depend on three-body correlations between the vortices as seen in the so-called ``vortex-glass state'' in La$_{1.9}$Sr$_{0.1}$CuO$_4$.\cite{Menon2006}\\
Therefore, in order to obtain accurate information on the superconducting-state parameters, it is preferable to complement transverse-field (TF) $\mu$SR measurements in the mixed state by investigations using other experimental techniques.
Here, we report on combined TF $\mu$SR and magnetization studies of the mixed state of a \lsco\ single crystal in a low magnetic field of $\mu_0H = 20$~mT applied parallel to the $c$ axis of the crystal. A detailed analysis shows that although the measurements are not susceptible to small variations in the order-parameter symmetry, the data can be overall consistently described by taking into account a single energy gap in the quasiparticle excitation spectrum with $d_{x^2-y^2}$ symmetry and a change in the sampling of the spatial field distributions by the muons due to vortex-pinning effects.

\section{Experimental details}
A cylindrical \lsco\ single crystal with a diameter of $5.5$~mm and a height of $10$~mm was used for the $\mu$SR studies. The crystal was grown by the travelling-solvent floating-zone technique similar to the one reported in Ref.~\onlinecite{Tanaka1989}. The subsequent characterization by Laue X-ray diffraction showed the $c$ axis pointing perpendicular to the cylinder axis. X-ray powder diffraction on a sample of the same batch indicated a single crystalline phase. The magnetization studies have been carried out in a Quantum Design 5T MPMS SQUID magnetometer on a small piece of approximate dimensions $5\times 4 \times 2$~mm$^3$ of the same batch with the $c$ axis along the shortest edge of the sample. The critical temperature $T_{\mathrm{c}}$ is about $37$~K (cf. Fig.~\ref{MvsTfc}). The magnetization measurements in a magnetic field in the range from a few millitesla up to several tesla applied parallel to the $c$ axis of the crystal were performed using different field-cooling (FC) procedures: (i) fast cooling from above $T_{\mathrm{c}}$ to $T=5$~K with a high rate of $\approx -20$~K/min and measuring during warming up the sample ($\mathrm{FFCW}$), (ii) slow cooling from above $T_{\mathrm{c}}$ to $T=5$~K with a low rate of $\approx -0.7$~K/min and measuring during this cooling ($\mathrm{SFCC}$), and (iii) measuring during warming up from the slowly field-cooled state obtained through (ii) ($\mathrm{SFCW}$).\\
The $\mu$SR experiments with magnetic fields up to $20$~mT applied along the $c$ axis were performed at the GPS spectrometer at the $\pi$M3 beam line at the Swiss Muon Source at the Paul Scherrer Institute.\cite{Abela1994} The used helium flow cryostat allows to cool the sample from $T=50$~K to $T=1.6$~K in about two minutes at the maximum cooling rate. It is therefore possible to do the $\mu$SR measurements using equivalent cooling procedures as for the magnetization studies. It should also be noted that the ratio of the sample dimensions parallel and perpendicular to the magnetic field has been similar for the magnetization and the $\mu$SR measurements. Thus, the demagnetization effects on the respective samples are similar for both studies and overall small due to the small absolute value of the susceptibility in the field-cooled mixed state.\\
For the $\mu$SR experiments spin-polarized positively charged muons with an energy of $\sim4$~MeV are implanted into the sample and thermalize. The spin ensemble interacts with its local environment until the muons decay ($\tau_{\mu}=2.197~\mu$s) and emit the formed positrons preferentially in the direction of the muon spin at the time of decay. Thus, by detecting the muons at their implantation time and the positrons after the decay the temporal evolution of the muon-spin polarization in a sample may be recorded and by that information about the local environment of the muons is obtained. In a static local magnetic field $B_{\mathrm{loc}}$ with a non-zero component perpendicular to the muon spins, these undergo a Larmor precession with a frequency $\omega = \gamma_{\mu} B_{\mathrm{loc}}$, where $\gamma_{\mu} = 2\pi\times135.54$~MHz/T is the gyromagnetic ratio of the muon. In a homogeneous transverse field the measured field distribution would therefore ideally be a Dirac $\delta$ function, whereas $P(B)$ has a finite width when spatially inhomogeneous fields are probed by the muons. In case the local fields within the mixed state of a type-II superconductor are sampled, this leads to a $\lambda$- and $\xi$-dependent characteristic asymmetric $P(B)$ with a ``high-field tail'' originating from contributions of muons stopping in and close to the vortex cores.\cite{Brandt1988} Further details concerning the analysis of these field distributions are introduced in Sec.~\ref{sec:musrdata}; more information on $\mu$SR techniques in general can be found in Ref.~\onlinecite{Yaouanc2011}.

\section{Magnetization studies of \lscohead}
\label{sec:magdata}
\begin{figure}[t]
\centering
\includegraphics[width=0.88\columnwidth]{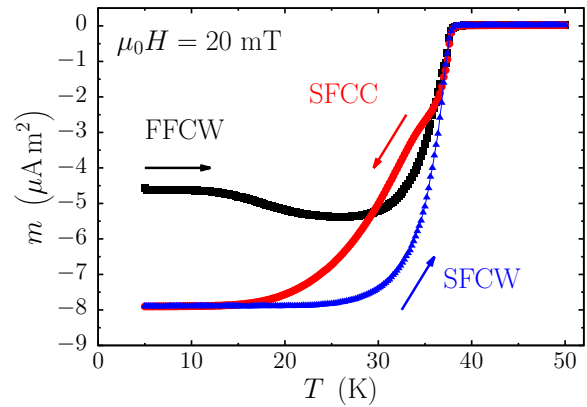}
\caption{(color online) Field-cooling magnetization curves of \lsco\ for $\mu_0H=20$~mT and $\mathbf{H}\parallel \mathbf{c}$ for the different cooling procedures described in the text.}
\label{MvsTfc}
\end{figure}

\begin{figure}[t]
\centering
\includegraphics[width=0.88\columnwidth]{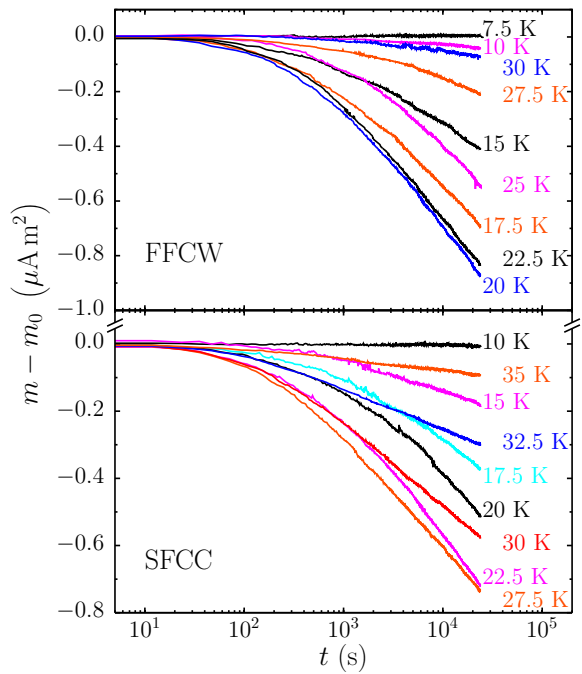}
\caption{(color online) Temporal evolution of the magnetic moment in an applied field of $20$~mT in the mixed state of \lsco\ obtained for the different cooling procedures. $m_0$ denotes the magnetic moment at $t=0$ defined for each temperature by data similar to those shown in Fig.~\ref{MvsTfc}.}
\label{Mrelaxation-20mT}
\end{figure}

\begin{figure}[t]
\centering
\includegraphics[width=0.88\columnwidth]{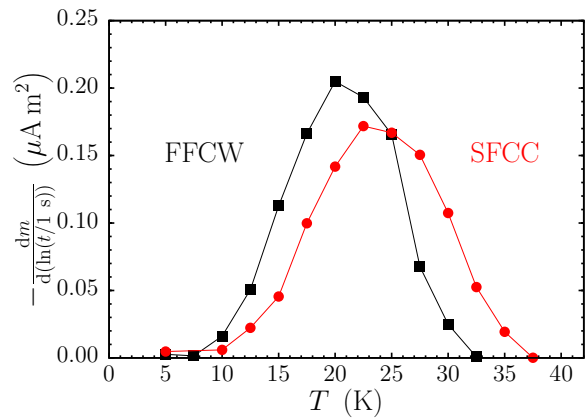}
\caption{(color online) Magnetic-moment relaxation rates in \lsco\ in $\mu_0H=20$~mT as a function of temperature for the two studied cooling procedures. The rates are determined from the data shown in Fig.~\ref{Mrelaxation-20mT} for $t>8100$~s.}
\label{RelRates-vs-T}
\end{figure}

The magnetization measurements in an applied magnetic field of $20$~mT are presented in Fig.~\ref{MvsTfc}. The overall diamagnetic signals exhibit certain peculiarities. While the slowly cooled $\mathrm{SFCW}$ data represent the magnetization curve of a rather equilibrated mixed state, the fast cooled $\mathrm{FFCW}$ measurement shows an initial low-temperature magnetization which deviates from the $\mathrm{SFCW}$ curve, reflecting a fairly undefined vortex configuration obtained by this cooling procedure. In contrast, the $\mathrm{SFCC}$ data show a hump below $T\approx35$~K.
The field-cooled magnetization of type-II superconductors during cooling (FCC) and during warming (FCW) were examined theoretically by Clem and Hao.\cite{Clem1993} They demonstrated that the important parameters in analyzing the irreversible behavior of the low-field dc susceptibilities are the critical current density, the sample dimensions, and the lower critical field $H_{\mathrm{c}1}$. However, the theoretical model developed in Ref.~\onlinecite{Clem1993} is subject to various simplifications and cannot be directly adapted to the curves presented in Fig.~\ref{MvsTfc}. Nevertheless, the detected hysteresis between the cooling and warming measurements is readily explained by differences in the flux expulsion and reentering in the presence of vortex pinning.\cite{Clem1993} In order to elucidate the origin of the distinct observations in the case that the sample is cooled very fast, the magnetization relaxation of the $\mathrm{FFCW}$ and $\mathrm{SFCC}$ mixed states was studied over a time of about six hours for various temperatures between $T=5$~K and $T_{\mathrm{c}}$ as depicted in Fig.~\ref{Mrelaxation-20mT}. It is seen that for temperatures below about $10$~K the vortex configuration hardly relaxes at all. For higher temperatures the magnetization returns to the ``equilibrium one'' with a maximum in the relaxation rate between about $20$~K and $25$~K which decreases again for temperatures above. Figure~\ref{RelRates-vs-T} shows the relaxation rate, defined as the slope of the measured magnetic moment in Fig.~\ref{Mrelaxation-20mT} for times $t>8100$~s. The vortex relaxation for $\mathrm{FFCW}$ and $\mathrm{SFCC}$ is similar for both studied situations. The small observed differences in the relaxation rates seem to be merely related to the differing initial conditions. Even though a fully quantitative analysis of the vortex relaxation is beyond the scope of this article, the low-field flux-line dynamics appears to be different in at least the three temperature regions described before. A similar behavior is found in \bscco\ where especially in a low magnetic field different vortex-pinning regimes have been identified.\cite{Nideroest1996} Yet, \lsco\ is far less anisotropic than \bscco\ and the more three-dimensional vortex structure is dominated by Josephson coupling rather than electromagnetic coupling. Thus, the pinning of individual ``vortex pancakes'' like in \bscco\ at low temperatures appears unlikely. However, the overall observed magnetization relaxation still suggests a crossover between distinct vortex-pinning scenarios also in \lsco.\\
It should be noted as well that the applied field of only $20$~mT at low temperatures is \emph{smaller} than $\mu_0H_{\mathrm{c}1}^{\parallel \mathbf{c}}$ in almost optimally doped \lsco.\cite{Naito1990} As can be seen in Fig.~\ref{MvsTfc} this does not affect the macroscopic ``equilibrium magnetization'', yet, the complete loss of dynamics for $T<10$~K might reflect the ``final freeze-in'' of the flux distribution at about the temperature where $\mu_0H_{\mathrm{c}1}^{\parallel \mathbf{c}} \approx 20$~mT.\cite{Clem1993}\\
Similar differences between FCC and FCW magnetization data have also been observed for other applied fields (with slightly shifted temperature intervals). However, previous $\mu$SR experiments on a single crystal of \lsco\cite{Khasanov2007} have shown that the field of $20$~mT is high enough to introduce a sufficient amount of vortices in the material which generate the characteristic magnetic-field distribution in the type-II superconductor. Therefore, the magnitude of the applied field has also been chosen to be $20$~mT for the present study. In the following it shall be investigated how the $\mu$SR results change for the various vortex configurations studied above.

\section{$\boldsymbol{\mu}$SR studies of \lscohead}
\label{sec:musrdata}
\begin{figure}[ht!]
\centering
\includegraphics[width=0.88\columnwidth]{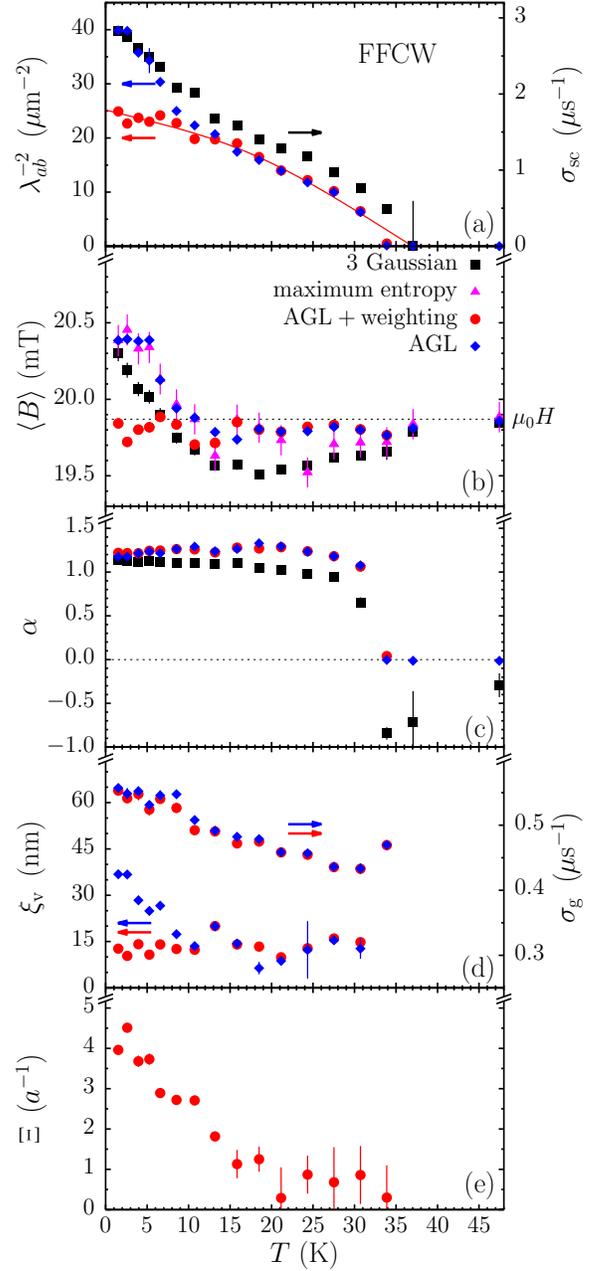}
\caption{(color online) Results of the analysis of the TF $\mu$SR data on \lsco: Temperature dependence of (a) $\lambda_{ab}^{-2}$ obtained from the AGL model and $\sigma_{\mathrm{sc}}$ from the ``3-Gaussian'' analysis; the solid red line is a fit to the semi-classical model of Ref.~\onlinecite{Chandrasekhar1993} using an order parameter with $d_{x^2-y^2}$ symmetry, (b) the probed mean field, (c) the skewness parameter $\alpha$, (d) the width of the Gaussian in Eq.~(\ref{asy-from-pB}) accounting for symmetric broadening of $P(B)$ within the AGL model and the effective vortex-core radius $\xi_{\mathrm{v}}$, and (e) the inverse width of the Lorentzian weighting function in Eq.~(\ref{Lorentzian}) given in units of the inverse inter-vortex distance of a fully ordered hexagonal flux-line lattice. $a = \sqrt{2\,\Phi_0/\left(\sqrt{3}\,\left\langle B\right\rangle\right)}$, where $\Phi_0 = h/(2\,e) = 2.07 \times 10^{-15}~\mathrm{T\,m}^{2}$ is the flux quantum. For further details, see text.}
\label{FCH}
\end{figure}

For the initial $\mu$SR studies the sample was cooled from above $T_{\mathrm{c}}$ to the lowest temperature in the applied field ${\mathbf{H}}\parallel \mathbf{c}$ at a cooling rate of $\approx -25$~K/min and the data were recorded while warming up the sample stepwise to $T>T_{\mathrm{c}}$. This temperature cycle corresponds to the $\mathrm{FFCW}$ procedure of the magnetization measurements. First, the resulting data were analyzed using a sum of three oscillating signals with Gaussian damping (``3-Gaussian'' method) which has been shown to be a reasonable approximation of the data if the probed field distribution originates from an ordered (or weakly distorted) vortex lattice.\cite{Maisuradze2009} In this case, the muon-decay asymmetry which is proportional to the spin polarization is modeled by
\begin{equation}
\mathcal{A}(t) = \sum_{i=1}^{3}A_i\exp\left(-\sigma_i^2t^2/2\right)\cos\left(\gamma_{\mu}B_it + \varphi\right).
\end{equation}
Here the $A_i$ are the partial asymmetries of the three signal contributions, $\sigma_i$ the respective Gaussian depolarization rates, and $B_i$ the average fields of each of the fractions. The $A_i$ are common parameters for all temperatures whereas the $\sigma_i$ and $B_i$ are temperature-dependent. $\varphi$ is the common initial phase of the muon spins with respect to the positron detector. The total decay asymmetry is $A=A_1+A_2+A_3$ and the second central moment of the corresponding field distribution is given by\cite{Weber1993}
\begin{equation}
\langle\Delta B^2\rangle = \frac{\sigma^2}{\gamma_{\mu}^2} = \frac{1}{A}\sum\limits_{i=1}^3A_i\left[\sigma_i^2/\gamma_{\mu}^2+\left(B_i-\langle B\rangle\right)^2\right],
\label{second-moment}
\end{equation}
where
\begin{equation}
\langle B\rangle = \frac{1}{A}\sum\limits_{i=1}^3A_iB_i
\end{equation}
is the first moment of the field distribution. Assuming the depolarization of the muon-spin ensemble is caused only by the inhomogeneous field distribution generated by the vortex lattice and random nuclear moments, then the contribution of the flux-line lattice to the second central moment is given by $\sigma_{\mathrm{sc}}^2=\sigma^2-\sigma_{0}^2$, where $\sigma_0 = 0.23(1)~\mu\mathrm{s}^{-1}$ is the muon-spin depolarization rate above $T_{\mathrm{c}}\approx 37$~K in the normal state of the material.
In order to further characterize the obtained field distributions also the dimensionless skewness parameter $\alpha\equiv\left\langle\Delta B^3\right\rangle^{1/3}/\left\langle\Delta B^2\right\rangle^{1/2}$ is calculated; it represents the lineshape asymmetry of $P(B)$.\cite{Lee1993, Yaouanc2011} Given the sum of Gaussian distributions, the third central moment can be written as
\begin{equation}
\langle\Delta B^3\rangle = \frac{1}{A}\sum\limits_{i=1}^3A_i\left(B_i-\left\langle B\right\rangle\right)\left[3\,\sigma_i^2/\gamma_{\mu}^2+\left(B_i-\langle B\rangle\right)^2\right].
\label{third-moment}
\end{equation}\\
For the final interpretations within the above model, the assumption\cite{Brandt1988II} $\sigma_{\mathrm{sc}}\propto\lambda_{ab}^{-2}$ ($\lambda_{ab}$ is the in-plane magnetic penetration depth) is essential. In order to test this assumption the data were also analyzed using an analytic Ginzburg-Landau (AGL) model.\cite{Hao1991, Yaouanc1997} Here the spatial distribution of the magnetic field in the vortex state of a superconductor is modeled by the Fourier series
\begin{equation}
B({\mathbf{r}}) = \langle B \rangle \sum\limits_{{\mathbf{K}}}\frac{f_{\infty}K_1\left[\dfrac{\xi_{\mathrm{v}}}{\lambda_{ab}}\left(f_{\infty}^2 + K^2\lambda_{ab}^2\right)^{\frac{1}{2}}\right]}{\left(f_{\infty}^2 + K^2\lambda_{ab}^2\right)^{\frac{1}{2}}K_1\left(\dfrac{\xi_{\mathrm{v}}}{\lambda_{ab}}f_{\infty}\right)}\exp\left(-\imath{\mathbf{K}}{\mathbf{r}}\right),
\label{AGL-model}
\end{equation}
where ${\mathbf{r}}=(x,y)$, ${\mathbf{K}}$ are the reciprocal lattice vectors of a two-dimensional hexagonal vortex lattice, $K=\vert\mathbf{K}\vert$, and $K_1$ is a modified Bessel function of the second kind. The parameter $f_{\infty}$ representing the suppression of the superconducting order parameter due to overlapping vortex cores has been set equal to $1$ (no suppression) in the analysis since the vortices are far apart in the low applied field used. The length $\xi_{\mathrm{v}}$ is the \emph{effective} vortex-core radius which is a variable parameter within this model. The field distribution $P(B)$ probed by the muons is obtained by random sampling of $B({\mathbf{r}})$ and finally the muon-decay asymmetry is given by
\begin{equation}
\mathcal{A}(t) =  A\exp\left(-\sigma_{\mathrm{g}}^2t^2/2\right)\int P(B)\cos\left(\gamma_{\mu} B t + \varphi\right) \mathrm{d}B.
\label{asy-from-pB}
\end{equation}
The Gaussian prefactor in Eq.~(\ref{asy-from-pB}) accounts for broadening of the field distribution by nuclear dipole fields as well as weak random pinning.\cite{Brandt1988} For the present data $\sigma_{\mathrm{g}}$ is growing continuously from $0.45~\mu\mathrm{s}^{-1}$ just below $T_{\mathrm{c}}$ to $0.55~\mu\mathrm{s}^{-1}$ at $T=1.6$~K (cf. Fig.~\ref{FCH}d).\\
The comparison of $\sigma_{\mathrm{sc}}$ obtained by the ``3-Gaussian'' analysis (black squares in Fig.~\ref{FCH}a) with $\lambda_{ab}^{-2}$ as determined by the AGL analysis (blue diamonds in Fig.~\ref{FCH}a) reveals that for this set of data both parameters cannot be simply scaled to match in the full temperature range, yet, both curves show pronounced changes in their curvatures around $T\approx13$~K.
The skewness parameter $\alpha$ is depicted in Fig.~\ref{FCH}c. Below $T\approx20$~K $\alpha$ is virtually constant; the small differences in the absolute value of $\alpha$ between the various models are merely the result of effectively cutting off high-field contributions in the ``3-Gaussian'' analysis, thus leading to an overall smaller $\alpha$ which is less susceptible to minor effects in the ``high-field tail'' of $P(B)$. One may note that the ``3-Gaussian'' analysis exhibits a strong drop and a sign change in $\alpha$ at $T=33(2)$~K. While such behavior usually is attributed to a melting of the vortex lattice,\cite{Lee1993, Aegerter1997} here the detailed temperature dependence of $\alpha$ is most likely an artifact of the fitting procedure involving three Gaussians for an almost symmetric field distribution slightly below $T_{\mathrm{c}}$. This view is supported by the fact that the irreversibility line in \lscox\ at this low applied magnetic field tends to be much closer to $T_{\mathrm{c}}$.\cite{Sasagawa2000, Gilardi2004}\\
However, it is most important to check the temperature dependence of $\langle B\rangle$ determined by the different analyses. Figure~\ref{FCH}b shows $\langle B\rangle$ for the above described models as well as its values obtained by a maximum-entropy approach\cite{Riseman2003} which has the advantage that it is \emph{not} tied to any model. For the maximum-entropy analysis an apodization with a Gaussian ($\sigma_{\mathrm{apod}} =1~\mu\mathrm{s}$) was applied to the asymmetry spectra. This leads to additional symmetric broadening of the determined field distribution but does not change $\langle B\rangle$. The various analyses all yield qualitatively the same results: while the mean field is slightly diamagnetically shifted and approximately constant for $T>13$~K, for $T\leqslant 13$~K $\langle B\rangle$ strongly \emph{rises} with decreasing temperature---even substantially above the applied field. Additionally, as shown in Fig.~\ref{FCH}d, in order to explain the data $\xi_{\mathrm{v}}$ obtained from the AGL fit would have to increase drastically in the same temperature range (cf. Fig.~\ref{FCH}d) whereas it assumes values between about $10$~nm and $15$~nm above $T\approx13$~K before it diverges close to $T_{\mathrm{c}}$.\\
Since the magnetization measurements have revealed a change in the vortex dynamics at $T\approx 10$~K and especially that the mixed state, generated in this initial $\mu$SR experiment, is far from being equilibrated it shall further be investigated if the increase in $\langle B\rangle$, $\xi_{\mathrm{v}}$, and the second central moment of the $\mu$SR field distributions at low temperatures is related to those observations. For this purpose, additional $\mu$SR measurements have been performed. In a first step the single crystal has been cooled slowly in the applied magnetic field from above $T_{\mathrm{c}}$ to finally $T=1.6$~K with a cooling rate of $-0.2$~K/min while the measurements have been done at intermediate stable temperature steps lasting about one hour each ($\mathrm{SFCC}$). Subsequently, the so generated ``equilibrium vortex state'' served as starting point for another series of measurements which have been conducted during the stepwise warming to above the critical temperature ($\mathrm{SFCW}$).
\begin{figure}[t]
\centering
\includegraphics[width=.8\columnwidth]{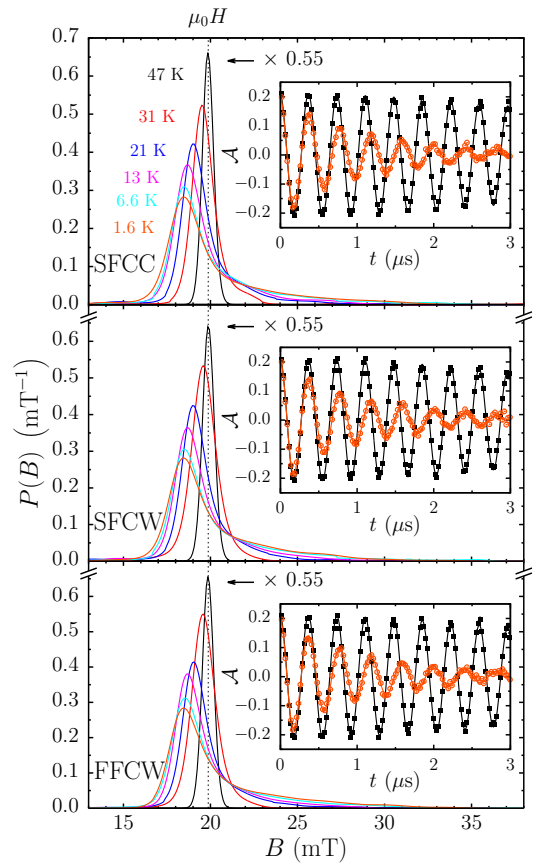}
\caption{(color online) Normalized magnetic field distributions for selected temperatures obtained by a maximum-entropy analysis of the TF $\mu$SR data on \lsco\ for the different cooling procedures. The insets show the corresponding asymmetry spectra for $T=47$~K (black squares) and $T=1.6$~K (orange circles), respectively. The solid lines in the insets are fits to the data using Eq.~(\ref{asy-from-pB}). For $T=1.6$~K, $P(B)$ is calculated using Eqs.~(\ref{AGL-model}) and~(\ref{Lorentzian}); for $T=47~\mathrm{K}>T_{\mathrm{c}}$, $P(B)=\delta\left(B-\mu_0H\right)$.}
\label{MaxEnt}
\end{figure}
\begin{figure}[ht!]
\centering
\includegraphics[width=0.88\columnwidth]{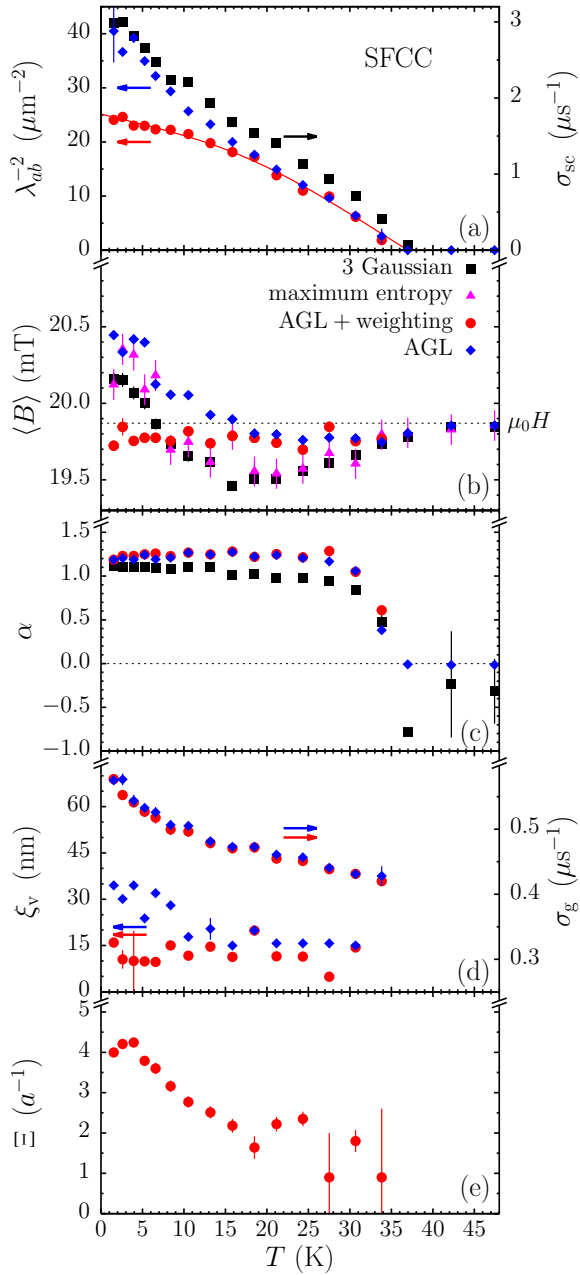}
\caption{(color online) Temperature dependence of the various model parameters for the $\mathrm{SFCC}$ TF $\mu$SR measurements of \lsco---analogous to Fig.~\ref{FCH}.}
\label{SCC}
\end{figure}
\begin{figure}[ht!]
\centering
\includegraphics[width=0.88\columnwidth]{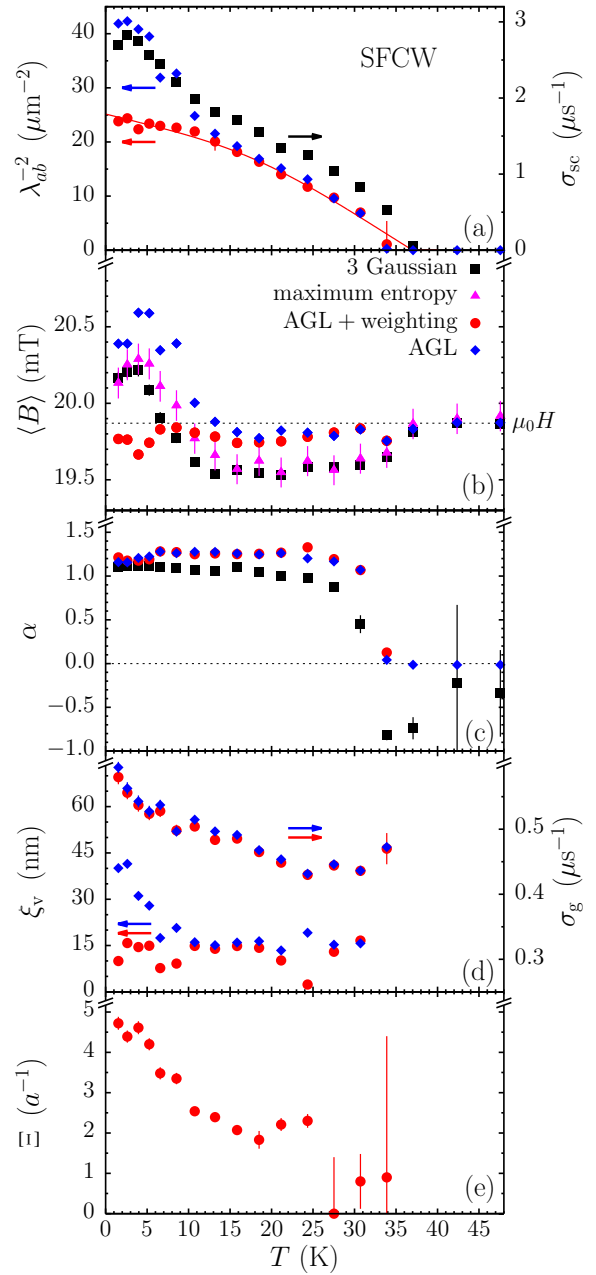}
\caption{(color online) Temperature dependence of the various model parameters for the $\mathrm{SFCW}$ TF $\mu$SR measurements of \lsco---analogous to Fig.~\ref{FCH}.}
\label{SCH}
\end{figure}
Figure~\ref{MaxEnt} shows a comparison between the static field distributions obtained by the maximum-entropy analysis of the TF $\mu$SR data as well as selected asymmetry spectra for the different cooling procedures---the differences are only marginal. The further analysis with the above described models confirms this finding (cf. Figs.~\ref{SCC}a--d and \ref{SCH}a--d). Also, the determined field distributions presented in Fig.~\ref{MaxEnt} agree well with the expectations where, e.\,g. given the model parameters obtained by the AGL analysis at the lowest temperature the ``high-field tail'' of $P(B)$ should range up to $32$~mT. The strikingly similar results for the different vortex configurations indicate that the $\mu$SR measurements at this small applied field mostly probe the local arrangements of the flux lines, rather than their long-range order. This is especially seen in the broadening parameter $\sigma_{\mathrm{g}}$ which originally had been introduced in the model to take into account nuclear dipolar broadening and disorder of the vortex lattice. The temperature dependence of this parameter is almost identical for the different cooling procedures, even though the ``vortex lattice'' should be much more disordered and essentially not clearly defined in the initial $\mathrm{FFCW}$ measurement, thus strongly suggesting that the broadening of the field distribution is not just related to vortex-lattice disorder.\\
The magnetization measurements do not yield any confirmation for an increased $\langle B \rangle$ due to the vortex lattice itself on the macroscopic scale---the applied field and the magnetization are constant below $T=10$~K (cf. Fig.~\ref{MvsTfc}). Hence, it is unlikely that the sudden increase in $\left\langle\Delta B^2\right\rangle$ below $T=13$~K---which is inevitably connected to the rising $\langle B \rangle$---is related to an increase in superfluid density. Moreover, $\lambda_{ab}(0)$ as obtained by the AGL model with the freely varying parameters is as low as $160$~nm which seems to be far too short for \lsco.\cite{Luke1997, Li1993} Furthermore, while the observation of the rather large effective vortex-core radius at $13~\mathrm{K}<T<T_{\mathrm{c}}$ is in agreement with previous studies of \lscox, where an expansion of the vortex cores has been found in a relatively low applied field,\cite{Kadono2004} the low-temperature values appear to be unreasonably large.\\

\section{Possible explanation for the obtained field distributions}

\begin{figure}[t]
\centering
\includegraphics[width=.88\columnwidth]{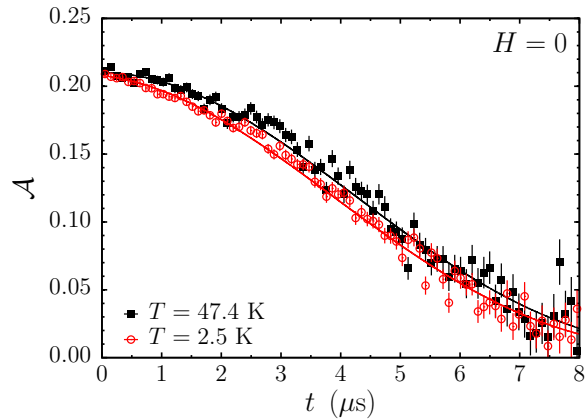}
\caption{(color online) Zero-field $\mu$SR asymmetry spectra obtained for \lsco\ at $T=47.4$~K and $T=2.5$~K. The low-temperature measurement shows a more exponential-like decay at small times and a slightly increased depolarization rate.}
\label{ZFmuSR}
\end{figure}

The origin of the increase in the probed mean field, the second central moment of $P(B)$, and the deduced effective vortex-core radius still remains to be clarified. As shown in Fig.~\ref{ZFmuSR} for the presently studied \lsco\ single crystal zero-field $\mu$SR measurements indicate a slightly enhanced spin depolarization at low temperatures---in accordance with the fact that the so-called ``cluster-spin-glass phase'' in \lscox\ might exist up to a doping level of $x=0.19$.\cite{Panagopoulos2002} Therefore, a small increase in the width of the probed field distribution at low temperatures ($T\lesssim3$~K) could be expected, however, static magnetic phases accounting for the dramatic rise in $\langle B\rangle$ for $T\lesssim13$~K can be excluded from these data.\\
The temperature dependence of $\langle B\rangle$ is strongly reminiscent of observations by $\mu$SR in \bscco;\cite{Harshman1991, Cubitt1993} there in applied fields of $0.3$~T to $0.4$~T at low temperatures the mean field probed by the muons also grows substantially with decreasing temperature whereas in the higher field of $1.5$~T the probed $\langle B\rangle$ equals the applied field. It has been argued that in low fields when the vortex motion ``freezes'' at low temperatures, the random sampling of the spatial field distribution of the vortex lattice is perturbed since pinning centers trap vortices and also offer possible muon stopping sites while at higher fields intervortex interactions dominate.\cite{Harshman1991} Of course, as already mentioned in Sec.~\ref{sec:magdata}, it has to be considered that the vortex-lattice topology is different in \lscox\ and the much more anisotropic \bscco\ system. The observed changes in the mean field in \bscco\ fall together with the transition from a ``vortex liquid'' to a ``vortex solid''\cite{Lee1993} whereas the melting of the vortex lattice in the present sample is much closer to $T_{\mathrm{c}}$ as discussed above already. Though, especially in relatively low magnetic fields \lscox\ shows a rich vortex-matter phase diagram where also in the solid phases the vortices are subject to thermal fluctuations.\cite{Divakar2004} Therefore, at high temperatures the vortices might ``hop'' thermally activated between different pinning centers which causes the muons only to be exposed to an average field [thus leading to a broadened but still asymmetric $P(B)$] while at low temperatures these thermal fluctuations are reduced and more muon spins precess in the high static fields close to the vortex cores which have a substantial effective size in the very low applied magnetic field. This could explain the observed positive shifts in the first moment and the second central moment of the measured field distribution. This view is supported by the magnetization data; even though the relevant time scales differ in the $\mu$SR and magnetization measurements, the observed almost cooling-cycle-independent change in the vortex dynamics around $T=10$~K in the magnetization data as shown in Figs.~\ref{Mrelaxation-20mT} and~\ref{RelRates-vs-T} seems to be correlated with the peculiar variations in the field distributions probed by the muons.\\
Very similar to the data discussed here---although in higher applied fields---Harshman~\emph{et al.} observed an inflection point in the temperature dependence of the second central moment of the field distribution measured by TF $\mu$SR in a single crystal of \ybco.\cite{Harshman2004} Also there, the increase of $\left\langle\Delta B^2\right\rangle$ at low $T$ was attributed to vortex-pinning effects which hence have been taken into account in the analysis. This analysis, however, does not provide an explanation for the increase in $\langle B\rangle$ as shown above. Therefore, instead of introducing pinning effects into the model for $\sigma_{\mathrm{sc}}$, in order to accomodate the possible enhanced correlations between the muons and the vortex cores in the analysis, the random sampling of the magnetic fields within the flux-line lattice is replaced by a phenomenological weighting function $w_{\Xi}$ enhancing the contributions of the fields close to the vortex cores to the resulting $P(B)$. The two-dimensional spatial field distribution calculated using Eq.~(\ref{AGL-model}) is weighted by the (not normalized) Lorentzian
\begin{equation}
w_{\Xi}(r_j) = \frac{1}{1+\Xi^2r_{j}^{2}},
\label{Lorentzian}
\end{equation}
where $r_j$ represents the distance to the center of the vortex core $j$ and $\Xi$ is the inverse width of the Lorentzian. For the calculation three to six neighboring vortex cores are taken into account. The obtained widths of the weighting function are depicted in Figs.~\ref{FCH}e, \ref{SCC}e, and \ref{SCH}e, respectively; above $T=20$~K where $P(B)$ is quite narrow, $\Xi$ is strongly correlated with $\sigma_{\mathrm{g}}$ but assumes generally small values, meaning that the fields are almost sampled randomly ($\Xi = 0$). For $T\leqslant 20$~K, $\Xi$ increases only modestly with decreasing temperature until $T=13$~K below which temperature the vortex-core contributions to the overall field distribution grow intensely---consistent with the changes in $P(B)$ observed during the previous analyses. On the other hand, $\langle B\rangle$ within this extended model remains essentially temperature-independent and slightly diamagnetically shifted with respect to the applied field (cf. Figs.~\ref{FCH}b, \ref{SCC}b, and \ref{SCH}b, red circles), overall consistent with the magnetization measurements. Due to the only small diamagnetic shift of $\langle B\rangle$ and the scattering of the determined mean-field values it cannot be finally decided if the observed difference in the absolute magnetic moment (cf. Fig.~\ref{MvsTfc}) for the employed cooling cycles is reflected in $\langle B\rangle$. (Please note that $\langle B\rangle$ in this case only refers to the mean field connected to the pure randomly sampled ordered vortex lattice, whereas the overall $\langle B\rangle$ of course is \emph{not} independent of temperature as discussed before.) Also, as seen in Figs.~\ref{FCH}d, \ref{SCC}d, and \ref{SCH}d the deduced $\xi_{\mathrm{v}}$ in this analysis is approximately constant as a function of temperature. The resulting temperature dependence of $\lambda_{ab}^{-2}$ is shown in Figs.~\ref{FCH}a, \ref{SCC}a, and \ref{SCH}a; it does \emph{not at all} show pronounced changes of the curvature around $T=13$~K and is approximately linear in $T$ at low temperatures. The full temperature dependence of $\lambda_{ab}^{-2}$ can be fairly described by the semi-classical model of Chandrasekhar and Einzel,\cite{Chandrasekhar1993, Prozorov2006} where the presented data are consistent with a weak-coupling $d_{x^2-y^2}$ order parameter with a maximum zero-temperature gap value $2\,\Delta_d(0) = 15.8(3)~\mathrm{meV} = 4.96(9)\,{k_{\mathrm{B}}T_{\mathrm{c}}}$ and a zero-temperature penetration depth $\lambda_{ab}(0) = 199(3)$~nm (solid red line in Figs.~\ref{FCH}a, \ref{SCC}a, and \ref{SCH}a). The cited uncertainties are of statistical nature only. Especially the value of $\lambda_{ab}(0)$ is subject to sizeable relative systematic errors of at least $5$~\% due to the phenomenological treatment of the vortex-pinning effects during the analysis. Nevertheless, the so determined value of $\lambda_{ab}(0)\approx200~\mathrm{nm}$ agrees well with data found in the literature~\cite{Luke1997, Li1993} while it appears to be estimated much too short if a randomly sampled vortex lattice was assumed as discussed at the end of Sec.~\ref{sec:musrdata}.\\
It should be noted that the presented explanation for the observed field distributions might not be the only possible one---yet, the introduced phenomenological model describes all obtained data consistently even though up to now the weight $\Xi$ cannot be fully related to the underlying quantities such as the pinning potential. Any other model would also have to be coherent and particularly it would have to provide an explanation for the temperature dependence of $\langle B\rangle$.\\
One such other possibility to explain the $\mu$SR data would be the presence of field- or vortex-induced antiferromagnetic order which would not be visible in the macroscopic magnetization measurements. This phenomenon has been reported to be present in cuprate high-temperature superconductors in applied fields of a few tesla by various techniques.\cite{Lake2001, Kakuyanagi2003, Miller2002} Yet, based on neutron-scattering results which indicate no ordered magnetic phase in low fields in \lscox\ with such high doping level as used in the present study, Sonier~{\emph{et~al.}} concluded from their $\mu$SR studies that only disordered field-induced magnetism is likely to occur in low applied fields.\cite{Sonier2007II} Therefore, the presence of an ordered magnetic state can be excluded for the presently discussed data.

\section{Discussion and Conclusions}

The TF $\mu$SR data presented in this work show a striking resemblance with data obtained earlier by Khasanov~\emph{et~al.} in $\mu$SR experiments on single crystals of cuprate high-temperature superconductors. Their data have been interpreted as evidence for two distinct energy gaps in the quasiparticle excitation spectrum, e.\,g. of \lsco,\cite{Khasanov2007} \ybco,\cite{Khasanov2007II} and YBa$_2$Cu$_4$O$_8$.\cite{Khasanov2008} Note that this is basically different from reports of slight deviations from a $d_{x^2-y^2}$ gap symmetry as measured, e.\,g. by phase-sensitive methods in \ybco.\cite{Kirtley2006} The conclusions of Ref.~\onlinecite{Khasanov2007} are mainly based on the observation of an inflection point in the temperature dependence of the second central moment of the magnetic field distribution $P(B)$ obtained by TF $\mu$SR measurements in the vortex state (like seen in Fig.~\ref{FCH}a), while the corresponding temperature dependence of the measured mean field has not been discussed at all. Our TF $\mu$SR data on \lsco\ are very similar, yet, the above presented analysis within the framework of conventional vortex-lattice-generated field distributions extended by vortex-pinning effects appears to describe the overall $\mu$SR and magnetization data in a coherent way. This is also in line with the suppression of the low-temperature increase of the second central moment of $P(B)$ in higher applied fields\cite{Khasanov2007} where inter-vortex interactions compete with vortex pinning and minimize its effects on the measured field distributions.\\
And, while it seems to be a reasonable approach to model two distinct gaps if an inflection point in $\lambda^{-2}(T)$ is found (e.\,g. $d_{x^2-y^2}+s$\cite{BussmannHolder2007} or $d_{x^2-y^2}+\imath{}d_{xy}$\cite{Valli2010}), the observation of the strong increase in the probed mean field $\langle B \rangle$ in the TF $\mu$SR measurements which is not observed in the magnetization measurements, renders it unlikely that the observed full second central moment of $P(B)$ is a true measure of $\lambda^{-2}$ at low temperatures. Moreover, it is pointed out, that the reported agreement between the obtained magnitudes of two-gap contributions from TF $\mu$SR and neutron crystal-field spectroscopy\cite{Furrer2007, Mueller2007} is probably fortuitous since fundamentally different models were employed to obtain those values. While in the analysis of the TF $\mu$SR data the contributions to the superfluid density were treated to be additive (\emph{after} the integrations over the Fermi surface),\cite{Khasanov2007, BussmannHolder2007} in the analysis of the neutron data the individual gaps are added together (\emph{before} the integration over the Fermi surface) yielding effectively \emph{one pseudogap} with an anisotropy $\vert\Delta_a/\Delta_b\vert\neq1$ and shifted nodes.\cite{Haefliger2006} Taking into account such an anisotropic \emph{superconducting gap} in the modeling of the superfluid density would result only in tiny overall changes from the $d_{x^2-y^2}$ scenario since firstly, only the modulus of the gap is relevant and secondly, the integration over the full Fermi surface\cite{Chandrasekhar1993} averages out small effects. Therefore, in general, a small gap anisotropy in the sense of Refs.~\onlinecite{Kirtley2006} or~\onlinecite{Haefliger2006} cannot be excluded, however, its determination from magnetic-penetration-depth data is rather difficult and indirect.\\

In conclusion, using low-field TF $\mu$SR in the mixed state of a \lsco\ single crystal a strong increase in the probed mean field and the second central moment of the measured field distribution is found at low temperatures. By combining $\mu$SR and magnetization measurements it is shown that these effects seem to be primarily related to vortex-pinning effects changing the sampling of the spatial field distribution by the muons and that the data can be consistently described by taking into account a single energy gap in the quasiparticle excitation spectrum with $d_{x^2-y^2}$ symmetry. However, small variations of the superconducting order parameter cannot be excluded on the basis of the presented $\mu$SR data. Given this assessment, the extrapolated magnetic penetration depth $\lambda_{ab}(0)\approx 200$~nm agrees well with reports of earlier experiments.\\
Furthermore, this article commemorates that TF $\mu$SR---in the way it has been employed for the present study---is first and foremost a very sensitive technique for the determination of local static or dynamic magnetic field distributions but the muon is neither a \emph{direct} probe of the superfluid density of a superconductor in the mixed state nor of its order-parameter symmetry as also has been realized earlier.\cite{Sonier2007III} Nevertheless, these parameters of a type-II superconductor can be deduced from TF $\mu$SR data in the vortex state if a coherent description of the determined field distributions is available.

\begin{acknowledgments}

The $\mu$SR measurements were performed at the Swiss Muon Source, Paul Scherrer Institute, Villigen, Switzerland. The $\mu$SR time spectra have been analyzed using the free software package \texttt{musrfit}\cite{Suter2010} mainly developed by A.~Suter and based on the CERN \texttt{ROOT} framework\cite{Brun1997} including the \texttt{Minuit} routines for function minimization.\cite{James1975} We thank M.~Bendele for his support during the $\mu$SR experiments. Helpful discussions with A.~Suter, A.~Furrer, E.~H. Brandt, and M.~Bendele are gratefully acknowledged. This work has been supported by the Swiss National Science Foundation and the NCCR MaNEP.

\end{acknowledgments}


\begin{thebibliography}{50}

\bibitem{Blundell1999} S.~J. Blundell, Contemp. Phys. {\bf{40}}, 175 (1999).

\bibitem{Sonier2000} J.~E. Sonier, J.~H. Brewer, and R.~F. Kiefl, Rev. Mod. Phys. {\bf{72}}, 769 (2000).

\bibitem{Blatter1994} G.~Blatter, M.~V. Feigel'man, V.~B.~Geshkenbein, A.~I. Larkin, and V.~M. Vinokur, Rev. Mod. Phys. {\bf{66}}, 1125 (1994).

\bibitem{Brandt1995} E.~H. Brandt, Rep. Prog. Phys. {\bf{58}}, 1465 (1995).

\bibitem{Brandt1988} E.~H. Brandt, J.~Low Temp. Phys. {\bf{73}}, 355 (1988).

\bibitem{Lee1993} S.~L. Lee, P.~Zimmermann, H.~Keller, M.~Warden, I.~M. Savi\'{c}, R.~Schauwecker, D.~Zech, R.~Cubitt, E.~M. Forgan, P.~H. Kes, T.~W. Li, A.~A. Menovsky, and Z.~Tarnawski, Phys. Rev. Lett. {\bf{71}}, 3862 (1993).

\bibitem{Menon2006} G.~I. Menon, A.~Drew, U.~K. Divakar, S.~L. Lee, R.~Gilardi, J.~Mesot, F.~Y. Ogrin, D.~Charalambous, E.~M. Forgan, N.~Momono, M.~Oda, C.~Dewhurst, and C.~Baines, Phys. Rev. Lett. {\bf{97}}, 177004 (2006).

\bibitem{Tanaka1989} Isao~Tanaka, Kenichi~Yamane, and Hironao~Kojima, J.~Cryst.~Growth {\bf{96}}, 711 (1989).

\bibitem{Abela1994} R.~Abela, C.~Baines, X.~Donath, D.~Herlach, D.~Maden, I.~D. Reid, D.~Renker, G.~Solt, and U.~Zimmermann, Hyperfine Interact. {\bf{87}}, 1105 (1994).

\bibitem{Yaouanc2011} A.~Yaouanc and P.~Dalmas de~R\'{e}otier, Muon Spin Rotation, Relaxation, and Resonance---Applications to Condensed Matter, Oxford University Press, 2011.

\bibitem{Clem1993} John~R. Clem and Zhidong~Hao, Phys. Rev.~B {\bf{48}}, 13774 (1993).

\bibitem{Nideroest1996} M.~Nider{\"o}st, A.~Suter, P.~Visani, A.~C. Mota, and G.~Blatter, Phys. Rev.~B {\bf{53}}, 9286 (1996).

\bibitem{Naito1990} M.~Naito, A.~Matsuda, K.~Kitazawa, S.~Kambe, I.~Tanaka, and H.~Kojima, Phys. Rev.~B {\bf{41}}, 4823 (1990).

\bibitem{Khasanov2007} R.~Khasanov, A.~Shengelaya, A.~Maisuradze, F.~La~Mattina, A.~Bussmann-Holder, H.~Keller, and K.~A. M{\"u}ller, Phys. Rev. Lett. {\bf{98}}, 057007 (2007).

\bibitem{Maisuradze2009} A.~Maisuradze, R.~Khasanov, A.~Shengelaya, and H.~Keller, J.~Phys. Condens.~Matter {\bf{21}}, 075701 (2009).

\bibitem{Weber1993} M.~Weber, A.~Amato, F.~N. Gygax, A.~Schenck, H.~Maletta, V.~N. Duginov, V.~G. Grebinnik, A.~B. Lazarev, V.~G. Olshevsky, V.~Yu. Pomjakushin, S.~N. Shilov, V.~A. Zhukov, B.~F. Kirillov, A.~V. Pirogov, A.~N. Ponomarev, V.~G. Storchak, S.~Kapusta, and J.~Bock, Phys. Rev.~B {\bf{48}}, 13022 (1993).

\bibitem{Brandt1988II} E.~H. Brandt, Phys. Rev.~B {\bf{37}}, 2349 (1988).

\bibitem{Hao1991} Zhidong Hao, John R. Clem, M. W. McElfresh, L. Civale, A. P. Malozemoff, and F. Holtzberg, Phys. Rev.~B {\bf{43}}, 2844 (1991).

\bibitem{Yaouanc1997} A.~Yaouanc, P.~Dalmas de R\'eotier, and E.~H. Brandt, Phys. Rev.~B {\bf{55}}, 11107 (1997).

\bibitem{Aegerter1997} C.~M. Aegerter and S.~L. Lee, Appl. Magn. Reson. {\bf{13}}, 75 (1997).

\bibitem{Sasagawa2000} T.~Sasagawa, Y.~Togawa, J.~Shimoyama, A.~Kapitulnik, K.~Kitazawa, and K.~Kishio, Phys. Rev.~B {\bf{61}}, 1610 (2000).

\bibitem{Gilardi2004} R.~Gilardi, A.~Hiess, N.~Momono, M.~Oda, M.~Ido, and J.~Mesot, Europhys. Lett. {\bf{66}}, 840 (2004).

\bibitem{Riseman2003} T.~M. Riseman and E.~M. Forgan, Physica B {\bf{326}}, 226 (2003).

\bibitem{Li1993} Qiang~Li, M.~Suenaga, T.~Kimura, and K.~Kishio, Phys.~Rev.~B {\bf{47}}, 2854 (1993).

\bibitem{Luke1997} G.~M. Luke, Y.~Fudamoto, K.~Kojima, M.~Larkin, J.~Merrin, B.~Nachumi, Y.~J. Uemura, J.~E. Sonier, T.~Ito, K.~Oka, M.~de~Andrade, M.~B. Maple, and S.~Uchida, Physica C {\bf{282}}, 1465 (1997).

\bibitem{Kadono2004} R.~Kadono, W.~Higemoto, A.~Koda, M.~I. Larkin, G.~M. Luke, A.~T. Savici, Y.~J. Uemura, K.~M. Kojima, T.~Okamoto, T.~Kakeshita, S.~Uchida, T.~Ito, K.~Oka, M.~Takigawa, M.~Ichioka, and K.~Machida, Phys. Rev. B {\bf{69}}, 104523 (2004).

\bibitem{Panagopoulos2002} C.~Panagopoulos, J.~L. Tallon, B.~D. Rainford, T.~Xiang, J.~R. Cooper, and C.~A. Scott, Phys. Rev.~B {\bf{66}}, 064501 (2002).

\bibitem{Harshman1991} D.~R. Harshman, R.~N. Kleiman, M.~Inui, G.~P. Espinosa, D.~B. Mitzi, A.~Kapitulnik, T.~Pfiz, and D.~Ll. Williams, Phys. Rev. Lett. {\bf{67}}, 3152 (1991).

\bibitem{Cubitt1993} R.~Cubitt, E.~M. Forgan, M.~Warden, S.~L. Lee, P.~Zimmermann, H.~Keller, I.~M. Savi\'{c}, P.~Wenk, D.~Zech, P.~H. Kes, T.~W. Li, A.~A. Menovsky, and Z.~Tarnawski, Physica C {\bf{213}}, 126 (1993).

\bibitem{Divakar2004} U.~Divakar, A.~J. Drew, S.~L. Lee, R.~Gilardi, J.~Mesot, F.~Y. Ogrin, D.~Charalambous, E.~M. Forgan, G.~I. Menon, N.~Momono, M.~Oda, C.~D. Dewhurst, and C.~Baines, Phys. Rev. Lett. {\bf{92}}, 237004 (2004).

\bibitem{Harshman2004} D.~R. Harshman, W.~J. Kossler, X.~Wan, A.~T. Fiory, A.~J. Greer, D.~R. Noakes, C.~E. Stronach, E.~Koster, and J.~D. Dow, Phys. Rev.~B {\bf{69}}, 174505 (2004).

\bibitem{Chandrasekhar1993} B.~S. Chandrasekhar and D.~Einzel, Ann. Phys. {\bf{505}}, 535 (1993).

\bibitem{Prozorov2006} Ruslan Prozorov and Russell~W. Giannetta, Supercond. Sci. Technol. {\bf{19}}, R41 (2006).

\bibitem{Lake2001} B.~Lake, G.~Aeppli, K.~N. Clausen, D.~F. McMorrow, K.~Lefmann, N.~E. Hussey, N.~Mangkorntong, M.~Nohara, H.~Takagi, T.~E. Mason, and A.~Schr\"{o}der, Science {\bf{291}}, 1759 (2001).

\bibitem{Kakuyanagi2003} K.~Kakuyanagi, K.~Kumagai, Y.~Matsuda, and M.~Hasegawa, Phys. Rev. Lett. {\bf{90}}, 197003 (2003).

\bibitem{Miller2002} R.~I. Miller, R.~F. Kiefl, J.~H. Brewer, J.~E. Sonier, J.~Chakhalian, S.~Dunsiger, G.~D. Morris, A.~N. Price, D.~A. Bonn, W.~H. Hardy, and R.~Liang, Phys. Rev. Lett. {\bf{88}}, 137002 (2002).

\bibitem{Sonier2007II} J.~E. Sonier, F.~D. Callaghan, Y.~Ando, R.~F. Kiefl, J.~H. Brewer, C.~V. Kaiser, V.~Pacradouni, S.~A. Sabok-Sayr, X.~F. Sun, S.~Komiya, W.~N. Hardy, D.~A. Bonn, and R.~Liang, Phys. Rev.~B {\bf{76}}, 064522 (2007).

\bibitem{Khasanov2007II} R.~Khasanov, S.~Str{\"a}ssle, D.~Di~Castro, T.~Masui, S.~Miyasaka, S.~Tajima, A.~Bussmann-Holder, and H.~Keller, Phys. Rev. Lett. {\bf{99}}, 237601 (2007).

\bibitem{Khasanov2008} R.~Khasanov, A.~Shengelaya, J.~Karpinski, A.~Bussmann-Holder, H.~Keller, and K.~A. M{\"u}ller, J.~Supercond. Nov. Magn. {\bf{21}}, 81 (2008).

\bibitem{Kirtley2006} J.~R.~Kirtley, C.~C.~Tsuei, A.~Ariando, C.~J.~M.~Verwijs, S.~Harkema, and H.~Hilgenkamp, Nat. Phys. {\bf{2}}, 190 (2006).

\bibitem{BussmannHolder2007} A.~Bussmann-Holder, R.~Khasanov, A.~Shengelaya, A.~Maisuradze, F.~La~Mattina, H.~Keller, and K.~A. M{\"u}ller, Europhys. Lett. {\bf{77}}, 27002 (2007).

\bibitem{Valli2010} A.~Valli, G.~Sangiovanni, M.~Capone, and C.~Di~Castro, Phys.~Rev.~B {\bf{82}}, 132504 (2010).

\bibitem{Furrer2007} A.~Furrer, J.~Supercond. Nov. Magn. {\bf{21}}, 1 (2008).

\bibitem{Mueller2007} K.~A. M{\"u}ller, J.~Phys. Condens. Matter {\bf{19}}, 251002 (2007).

\bibitem{Haefliger2006} P.~S. H{\"a}fliger, A.~Podlesnyak, K.~Conder, E.~Pomjakushina, and A.~Furrer, Phys. Rev. B {\bf{74}}, 184520 (2006).

\bibitem{Sonier2007III} J.~E. Sonier, Rep. Prog. Phys. {\bf{70}}, 1717 (2007).

\bibitem{Suter2010} A.~Suter and B.~M. Wojek, accepted for publication in Physics Procedia.\\
See also http://lmu.web.psi.ch/facilities/software/musrfit/\\ technical/index.html

\bibitem{Brun1997} R.~Brun and F.~Rademakers, Nucl. Instrum. Methods Phys. Res., Sect. A {\bf{389}}, 81 (1997).\\
See also http://root.cern.ch/

\bibitem{James1975} F.~James and M.~Roos, Comput. Phys. Commun. {\bf{10}}, 343 (1975).

\end{thebibliography}
\end{document}